\documentclass[12pt,thmsa]{article}
\usepackage{amsmath}
\usepackage{sw20lart}

\input{tcilatex}

\begin{document}

\title{Momentum of electromagnetic fields, speed of light in moving media,
and the photon mass}
\author{G. Spavieri \\
\textit{Centro de F\'{\i}sica Fundamental, Facultad de Ciencias, }\\
\textit{Universidad de Los Andes, } \textit{M\'{e}rida, 5101-Venezuela%
\thanks{%
spavieri@ula.ve} } \and G. T. Gillies \\
\textit{Department of Physics, University of Virginia}\\
\textit{Charlottesville, VA 22901-4714 USA}}
\date{}
\maketitle

\begin{abstract}
In both the equations for matter and light wave propagation, the momentum of
the electromagnetic fields $\mathbf{P}_{e}$ reflects the relevant em
interaction. As a review of some applications of wave propagation
properties, an {optical experiment which tests} the speed of light in moving
rarefied gases is described. Moreover, $\mathbf{P}_{e}$ is also the link to
the unitary vision of the quantum effects of the Aharonov-Bohm (AB) type,
which provide a useful quantum approach for the limit of the photon mass $%
m_{ph}$. A bench-top experiment based on effects of the AB type that exploit
new interferometric techniques, is foreseen to yield the limit $m_{ph}\simeq
10^{-54}g$, a value that improves upon the results achieved with other
approaches.

\end{abstract}

\section{Introduction}

The existing formal analogy between the wave equation for light in moving
media and that for charged matter waves has been described by Spavieri and
Gillies \cite{sg} in the context of a proposed optical experiment for light
propagation in transparent moving media, previously discussed by several
authors \cite{han}. The link between the two wave equations is the \textit{%
interaction }electromagnetic (em) momentum $\mathbf{P}_{e}$, which has
attracted physicists' attention because it arises in different scenarios of
modern physics involving em interactions. The first of these scenarios is
that of light propagation in slowly moving media \cite{han}, \cite{sg}, and
another involves quantum nonlocal effects of the Aharonov-Bohm (AB) type 
\cite{AB} and their unitary view \cite{s}. More commonly, the interaction%
\textit{\ }em momentum $\mathbf{P}_{e}$ appears as a nonvanishing quantity
in em experiments involving "open" or convection currents, while $\mathbf{P}%
_{e}$ vanishes in the common em experiments or interactions with closed
currents or circuits \cite{sg}, \cite{sgo}.

The main purpose of this article is to review the recent advances of physics
involving the em momentum $\mathbf{P}_{e}$ and its role in the proposal of
new tests or in making other advances, such as setting a new limit on the
photon mass. The arguments presented here are also been described in our
companion paper of Ref. \cite{sg1}.

In the field of electromagnetism, a growing number of articles questioning
the standard interpretation of special relativity have appeared \cite{1}-%
\cite{c}. Some of the authors of Refs. \cite{1} and \cite{4} adhere to a
point of view that assumes the existence of a preferred frame, similar to
the historical works of Lorentz and Poincar\'{e}. {\ It has been argued that
these different formulations of Special Relativity are truly compatible only
in vacuum, as differences may appear when light propagates in transparent
moving media}. Thus, Consoli and Costanzo \cite{c}, Cahill and Kitto \cite%
{ck}, and Guerra and de Abreu \cite{4}, point out that, for the experiments
of the Michelson--Morley type, which are often said to have given a
null-result, this is not the case and cite the famous work by Miller \cite%
{mil}. The claim of these authors is that the available data point towards a
consistency of non-null results when the interferometer is operated in the
\textquotedblleft gas-mode\textquotedblright , corresponding to light
propagating through a gas \cite{c} (as in the case of air or helium, for
instance, even in modern maser versions of optical tests).

Furthermore, the only tests involving "open" or convection currents, so far
historically performed, have been reconsidered by Indorato and Masotto \cite%
{r9} who points out that these experiments are not completely reliable and
may be inconclusive \cite{sg}. As a response to this, physicists have
recently proposed experiments about those predictions of the theory that
have not been fully tested, or they have formulated untested assumptions
that differ from the standard interpretation of Special Relativity \cite{sg}%
, \cite{sgo}, \cite{4}, \cite{c}.

The interesting point is that all the above-mentioned scenarios and
polemical hypotheses are linked to the interaction em momentum. Therefore,
throughout this article we highlight the role of $\mathbf{P}_{e}$ in each
one of these scenarios.

\section{Wave equations for matter and light waves}

To elucidate the role of em momentum in modern physics, we start by
considering the wave equations for matter and light waves and show how the
interaction term $\mathbf{Q}$ of these equations is related to $\mathbf{P}%
_{e}$ \cite{Sem}. In general, with $T_{ik}^{M}$ the Maxwell stress-tensor,
the covariant description of the em momentum leads to the four-vector em
momentum $P_{e}^{\alpha }$ expressed as 
\begin{equation}
P_{e}^{i}\,c=\gamma \int (c\mathbf{g}+T_{ik}^{M}\beta ^{i})d^{3}\sigma
\;\;\;\;\;\;\;\;\;\;\;\;cP_{e}^{0}=\gamma \int (u_{em}-\mathbf{v\cdot g}%
)d^{3}\sigma  \label{emm}
\end{equation}%
where\textbf{\ }$\beta =v/c$, and the em energy and momentum are evaluated
in a special frame $K^{(0)}$ moving with velocity $\mathbf{v}$ with respect
to the laboratory frame. Here, $u_{em}$ is the energy density and $\mathbf{S}%
=\mathbf{g}c$ is the energy flux or flow.

The analogy between the wave equation for light in moving media and that for
charged matter waves has been pointed out by Hannay \cite{han} and later
addressed by Cook, Fearn, and Milonni \cite{han} who have suggested that
light propagation at a fluid vortex is analogous to the Aharonov-Bohm (AB)
effect, where charged matter waves (electrons) encircle a localized magnetic
flux \cite{AB}. Generally, in quantum effects of the AB type \cite{AB}-\cite%
{s} matter waves undergo an em interaction as if they were propagating in a
flow of em origin that acts as a moving medium \cite{s} and modifies the
wave velocity. This analogy has led to the formulation of the so-called 
\textit{magnetic model of light propagation} \cite{han}, \cite{sg}.

According to Fresnel \cite{fre}, light waves propagating in a transparent,
incompressible moving medium with uniform refraction index $n$, are dragged
by the medium and develop an interference structure that depends on the
velocity $\mathbf{u}$ of the fluid ($u<<c$). At the time of Fresnel the
preferred inertial frame was that at rest with the so-called ether, which
here may be taken to coincide with the laboratory frame. The speed achieved
in the ether frame is 
\begin{equation}
v=\frac{c}{n}+(1-\frac{1}{n^{2}})\,u  \label{vf}
\end{equation}%
as later corroborated by Fizeau \cite{fre}. Because of the formal analogy
between the wave equation for light in slowly moving media and the Schr\"{o}%
dinger equation for charged matter waves in the presence of the external
vector potential $\mathbf{A}$ (i.e., the magnetic Aharonov-Bohm effect),
both equations contain a term that is generically referred to as the
interaction momentum $\mathbf{Q}$. Thus, the Schr\"{o}dinger equation for
quantum effects of the AB type (with $\hbar =1$) \cite{s} and the wave
equation for light in moving media can be written \cite{han}, \cite{sg} as 
\begin{equation}
(-i\mathbf{\nabla }-\mathbf{Q})^{2}\Psi =p^{2}\Psi .  \label{ab1}
\end{equation}

Eq.(\ref{ab1}) describes matter waves if the momentum $p$ is that of a
material particle, while, if $p$ is taken to be the momentum $\hbar k$ of
light (in units of $\hbar =1$), Eq.(\ref{ab1}) describes light waves.

a) All the effects of the AB type discussed in the literature \cite{AB}-\cite%
{s} can be described by Eq.(\ref{ab1}), provided that the interaction
momentum $\mathbf{Q}$ is related \cite{s}, \cite{Sem} to $\mathbf{P}_{e}$,
the momentum of the em fields. The AB term $\mathbf{Q}=(e/c)\mathbf{A}$ of
the magnetic AB effect is obtained by taking $\mathbf{Q}=\mathbf{P}_{e}=%
\frac{1}{4\pi c}\int (\mathbf{E\times B})d^{3}\mathbf{x}^{\prime }$ where $%
\mathbf{E}$ is the electric field of the charge and $\mathbf{B}$ the
magnetic field of the solenoid. A general proof that this result holds in
the \textit{natural }Coulomb gauge, has been given by several authors \cite%
{boy}. For these quantum effects, the solution to Eq. (\ref{ab1}) is given
by the matter wave function 
\begin{equation}
\Psi =e^{i\phi }\Psi _{0}=e^{i\int \mathbf{Q\cdot }d\mathbf{x}}\,\Psi
_{0}=e^{i\int \mathbf{Q\cdot }d\mathbf{x}}\,e^{i(\mathbf{p\cdot x}-Et\mathbf{%
)}}\,\mathcal{A}  \label{psi}
\end{equation}%
where $\Psi _{0}$ solves the Schr\"{o}dinger equation with $\mathbf{Q}=0$.

b) Calculations of the quantity $\mathbf{Q}=\mathbf{P}_{e}$ (\ref{emm}) for
light in slowly moving media show \cite{Sem} that the interaction term
yields the Fresnel-Fizeau momentum \cite{sg} 
\begin{equation}
\mathbf{Q}=-\frac{\omega }{c^{2}}(n^{2}\mathbf{-}1)\mathbf{u,}  \label{qhk}
\end{equation}%
and that a solution of the type described in (\ref{psi}) may assume the
forms 
\begin{equation}
\Psi =e^{i\phi }\Psi _{0}=\,e^{i\int \mathbf{Q\cdot }d\mathbf{x}}e^{i\int (%
\mathbf{k\cdot }d\mathbf{x}-\omega \,dt\mathbf{)}}\mathcal{A};\;\;\;\Psi
=\,e^{i\int (\mathbf{K}(\mathbf{x})\mathbf{\cdot }d\mathbf{x}-\omega \,dt%
\mathbf{)}}\mathcal{A}  \label{psi2}
\end{equation}%
where $\mathbf{k}$ and $\mathbf{K}(\mathbf{x})$ are wave vectors, $\omega
=k\,c/n$ the angular frequency, and $n$ the index of refraction, while $\Psi
_{0}$ solves Eq.(\ref{ab1}) with $\mathbf{Q}=\mathbf{u}=0$.

The fact that the interaction momentum $\mathbf{Q}$ is related to $\mathbf{P}%
_{e}$ \cite{s}, \cite{Sem} for both matter waves of effects of the AB type 
\cite{s} and light waves in moving media \cite{Sem}, definitely reinforces
the existing analogy between the two wave equations. Two theoretical
possibilities arise \cite{sg}:

- By incorporating the phase $\phi $ in the term $\int \mathbf{K}(\mathbf{x})%
\mathbf{\cdot }d\mathbf{x}$, the last expression on the rhs of Eq.(\ref{psi2}%
) keeps the usual invariant form of the solution as required by special
relativity and one finds \cite{Sem} for the speed of light the result $%
\mathbf{v}=(c/n)\widehat{\mathbf{c}}+(1\mathbf{-}1/n^{2})\mathbf{u}=(c/n)%
\widehat{\mathbf{c}}-\mathbf{Q}(c^{2}/n^{2}\omega )$ in agreement with Eq.(%
\ref{vf}) and Special Relativity.

- Maintaining instead the analogy with the AB effect, the solution can be
chosen to be represented by the first term of Eq.(\ref{psi2}), $\Psi
=e^{i\phi }\Psi _{0}$. In this case, the phase velocity changes but the
speed of light (the particle, or photon) may not change \cite{sg}. This
result is in total agreement with the analogous result for the AB effect
where $\mathbf{Q}=(e/c)\mathbf{A}$ and the particle speed is left unchanged
by the interaction with the vector potential $\mathbf{A}$.

The established relation (\ref{qhk}) will be used in the next sections to
tentatively express in a quantitative way the hypothesis of Consoli and
Costanzo \cite{c} referring to $v$, the speed of light in a moving rarefied
media. With a quantitative expression for $v$ it is then possible to
formulate a dedicated experiment that tests Consoli and Costanzo's
hypothesis.

\subsection{Propagation of em waves in rarefied moving media}

Duffy \cite{duf} has noted that the concept of an ether-like preferred frame
has always incited controversy, even in modern scientific investigations
aimed at exploring the less understood aspects of relativity theory. Within
this scenario, Consoli and Costanzo \cite{c}, Cahill and Kitto \cite{ck},
and Guerra and de Abreu \cite{4}, after a re-analysis of the optical
experiments of the Michelson--Morley type, claim that the available data
point towards a consistency of non-null results when light in the arms of
the interferometer propagates in a rarefied gas, like the cases of air at
normal pressure and temperature. The possibility of maintaining the
existence of a preferred frame\textbf{, }and parallel interests in the
Michelson-Morley, Trouton-Noble and related effects, arises because the
coordinate transformation used, the Tangherlini transformations \cite{tan}
foresee the same length contraction and time dilation of the Lorentz
transformations. However, they contain an arbitrariness in the determination
of the time synchronization parameter, with the consequence that there are
quantities which eventually cannot be measured, such as the one-way speed of
light, its measured value depending on the synchronization procedure adopted 
\cite{tan}. Different synchronization procedures are possible \cite{1}-\cite%
{c}, fully compatible with Einstein's relativity in practice, but with very
different assertions in fundamental and philosophical terms.

The original important assumption made by Consoli et al. to corroborate
their claims of a non-null result and open a window for the possible
existence of a preferred frame, is that light in a moving rarefied gas of
refractive index $n$ very close to $1$ propagates with speed $c/n$ ,
isotropically, in the preferred frame, as if the medium were not moving.
Obviously, this hypothesis is in contrast with special relativity that
foresees the speed (\ref{vf}), but it is not ruled out by the known optical
tests. Thus, this assumption needs justification and experimental
corroboration.

In the following, we explore possible modifications of the form of the
present Fresnel-Fizeau momentum when the moving medium is composed of
rarefied gas. It is not unconceivable that the effectiveness of the light
delay mechanism in a compact moving medium differs, and perhaps even
substantially so, from that of a non-compact moving medium, such as a
rarefied gas, even if they have the same index $n$. As an \textit{ad hoc}
hypothesis or a tentative model of a light delay mechanism, it has been
supposed \cite{sgv} that its effectiveness $e_{f}$ arises from the relative
spatial extension $V_{i}$ of the interaction em momentum $\mathbf{Q}(\mathbf{%
u})$ with respect to the extension $V$ of the total em momentum. Introducing
then the ratio $e_{f}=V_{i}/V$, the effective em interaction momentum, to be
used in determining the speed of light in a moving media, will be assumed to
be given by the effective Fresnel-Fizeau term $\,e_{f}\,\mathbf{Q}%
=\,(V_{i}/V)\,\mathbf{Q}$, while the resulting velocity of light in moving
rarefied media is 
\begin{equation}
\mathbf{v}=\frac{c}{n}\widehat{\mathbf{c}}-\frac{c^{2}}{n^{2}\omega }e_{f}\,%
\mathbf{Q}=\frac{c}{n}\widehat{\mathbf{c}}+e_{f}\,(1-\frac{1}{n^{2}})\,%
\mathbf{u}.  \label{cr}
\end{equation}%
The hypothesis of Consoli et al. of the speed $c/n$ in the preferred frame
for moving rarefied gases, will be justified by our model if $e_{f}=V_{i}/V$
turns out to be very small and, in this case, negligible. Calculations
leading to a rough estimate of $V_{i}/V$ for air at room temperature yield 
\cite{sgv} $e_{f}=N_{a}(a^{3}/R^{3})\,22.9=6.1\times 10^{-3}$, which indeed
can be neglected. Thus, our model foresees that the speed of light in moving
media is actually not $c/n$ but, quantitatively, the changes found do not
alter significantly the basic hypothesis and resulting analysis by Consoli
et al. \cite{c}, \cite{ck} and Guerra et al. \cite{4}.

\section{\protect\bigskip Optical test in the first order in $v/c$}

The main consequence is that, with the present hypothesis of negligible
drag-like effect for moving rarefied gases, ether drift experiments of the
order $v/c$ become meaningful again. Let us consider for example the
following experiment which is a variant of the Mascart and Jamin experiment
of 1874 \cite{mj}.

A ray of light travels from point A to point B of a segment A--===--B
representing an optical interferometer. The original ray is split into two
rays at A, which propagate separately through the two arms (1 and 2) of the
interferometer. The rays recombine then at B where the interference pattern
is observed. The arms 1 and 2 are made of a transparent rarefied gases or
materials with indices of refraction $n_{1}$ and $n_{2}$ and wherein the
speeds are $c/n_{1}$ and $c/n_{2}$ in the preferred frame, respectively, in
agreement with Consoli's et al. hypothesis \cite{c} of the velocity
expression (\ref{cr}) with $e_{f}=0$. The laboratory frame with the
interferometer and the rarefied gas is moving with speed $u$ with respect to
the preferred frame. {We could be using the expressions for the speed in the
moving laboratory frame resulting from\ the Tangherlini transformation,
which can be found in \cite{tan}, \cite{4}. The calculation can also be done
using the standard velocity addition from the Lorentz transformation, 
\textit{i.e.}, using the definition of \textit{Einstein speed} as detailed
in \cite{4}. Both approaches yield the same result. The speed of light in
arm 1 in the frame of the interferometer, moving with speed }$u${\ with
respect to the preferred frame, is respectively} 
\begin{equation}
w_{1}=\frac{c/n_{1}-u}{1-u^{2}/c^{2}}\;\;\text{ or}\;\;\text{ }w_{1}=\frac{%
c/n_{1}-u}{1-u/(c\,n_{1})},  \label{spe}
\end{equation}%
and analogously for $w_{2}$.{\ If $L$ is the length of the arms, the time
delay, or optical path difference, for the two rays yields, in the first
order in }$u/c$, 
\begin{equation}
\Delta t(0^{o})=L(\frac{1}{w_{1}}-\frac{1}{w_{2}})\simeq \frac{L}{c}%
(n_{1}-n_{2})[1+\frac{u\,}{c}(n_{1}+n_{2})].  \label{zer}
\end{equation}%
In order to observe a fringe shift,{\ the interferometer needs to be
rotated, typically by 90 or 180 degrees. The time delay for 180 degrees is
the same of Eq.(\ref{zer}) with }$u$ replaced by $-u$. {The observable
fringe shift upon rotation of the interferometer does not vanish in the
first order in $u/c$ and is related to the time delay variation} 
\begin{equation}
\delta t=\Delta t(0^{o})-\Delta t(180^{o})\simeq 2\frac{u}{c}%
(n_{1}^{2}-n_{2}^{2})\frac{L}{c}.
\end{equation}

Choosing two media with different refractive index such that $%
n_{1}^{2}-n_{2}^{2}$ is not too small ($>10^{-3})$, the resulting fringe
shift should be easily observable if the preferred frame exists and its
speed $u$ is not too small. Knowing the sensitivity of the apparatus, one
could set the lower limit of the observable preferred speed $u$.
Interferometers, used in advanced Michelson-Morley's type of experiments,
could detect a speed $u$ as small as $1km/s$ (a few $m/s$ for He-Ne maser
tests). Thus, this optical experiment, in passing from second order ($%
u^{2}/c^{2}$) to first order tests, should be able to improve the range of
detectability of $u$ by a factor $(c/u)(n_{1}^{2}-n_{2}^{2})\simeq 3\times
10^{5}\times 10^{-3}=3\times 10^{2}$, i.e., detect with the same
interferometer speeds $3\times 10^{2}$ smaller.

New, more refined versions of the Michelson-Morley type of experiment
(including the tests using He-Ne masers.) are not suitable to test the
hypothesis of Consoli et al. \cite{c} because of the relatively low
sensitivity of these experimental approaches for rarefied gases. However, as
shown above, an optical test in the first order in $v/c$ becomes meaningful
in this case and can provide important advantages over the second order
experiments of the Michelson-Morley type.

\section{Effects of the Aharonov-Bohm type and the photon mass}

We have shown in the previous sections that all the effects of the AB type
can be described in a unified way by the wave equation (\ref{ab1}) where,
for each one of the effects, the quantity $\mathbf{Q}$ represents the em
interaction momentum (\ref{emm}). Both the interaction energy and momentum
appear in the expression of the phase of the quantum wave function. Through
the phenomenon of interference, phase variations can be measured and the
observable quantity can be related to variations of the interaction em
momentum or energy. In the following sections we show how the photon mass
can be determined by measuring its effect on the observable phase variation
via the related changes of em momentum or energy.

The possibility that the photon possesses a finite mass and its physical
implications have been discussed theoretically and investigated
experimentally by several researchers \cite{wfh}, \cite{lu}. Originally, the
finite photon mass $m_{\gamma }$ (measured in \textit{centimeters}$^{-1}$)
has been related to the range of validity of Coulomb law \cite{wfh}. If $%
m_{\gamma }\neq 0$ this law is modified by the Yukawa potential $%
U(r)=e^{-m_{\gamma }\,r}/r$, with $m_{\gamma }^{-1}=\hbar /m_{ph}c=\lambda
_{C}/2\pi $ where $m_{ph}$ is expressed in \textit{grams} and $\lambda _{C}$
is the Compton wavelength of the photon.

There are direct and indirect tests for the photon mass, most of them based
on classical approaches. Recalling some of the classical tests, we mention
the results of Williams, Faller and Hill \cite{wfh} yielding the range of
the photon rest mass $m_{\gamma }^{-1}>3\times 10^{9}cm$, and of Luo, Tu,
Hu, and Luan \cite{lu} yielding the range $m_{\gamma }^{-1}>1.66\times
10^{13}cm$ and corresponding photon mass $m_{ph}<$ $2.1\times 10^{-51}g$.

Several conjectures related to the Aharonov-Bohm (AB) effect have been
developed assuming electromagnetic interaction of fields of infinite range,
i.e., zero photon mass. The possibility that any associated effects become
manifest within the context of finite-range electrodynamics has been
discussed by Boulware and Deser (BD) \cite{bd}. In their approach, BD
consider the coupling of the photon mass $m_{\gamma }$, as predicted by the
Proca equation $\partial _{\nu }F^{\mu \nu }+m_{\gamma }^{2}A^{\mu }=J^{\mu
} $, and calculate the resulting magnetic field$\;\mathbf{B}=\mathbf{B}_{0}+%
\widehat{\mathbf{k}}\,m_{\gamma }^{2}\,\Pi (\rho )$, that might be used in a
test of the AB effect. Because of the extra mass-dependent term, BD obtained
a nontrivial limit on the range of the transverse photon from a table-top
experiment yielding $m_{\gamma }^{-1}>1.4\times 10^{7}cm$.

After the AB effect, other quantum effects of this type have been developed,
such as those associated with neutral particles that have an intrinsic
magnetic \cite{sdual} or electric dipole moment \cite{s}, and those with
particles possessing opposite electromagnetic properties, such as opposite
dipole moments or charges \cite{s}, \cite{sep}-\cite{dow}. The impact of
some of these new effects on the photon mass has been studied by Spavieri
and Rodriguez (SR) \cite{SR}.

Based on theoretical arguments of gauge invariance, SR point out that, in
analogy with the AC effect for a coherent superposition of beams of magnetic
dipoles of opposite magnetic moments $\pm \mu $ \cite{sang} and the effect
for electric dipoles of opposite moments $\pm d$ \cite{dow}, the Spavieri
effect \cite{sep} of the AB type for a coherent superposition of beams of
charged particles with opposite charge state $\pm q$ is theoretically
feasible. Using this effect, SR evaluate its relevance in eventually
determining a bound for the photon mass $m_{ph}$. SR consider a coherent
superposition of beams of charged particles with opposite charge state $\pm
q $ passing near a huge superconducting cyclotron. The $\pm $ charges feel
the effect of the vector potential $\mathbf{A}$ created by the intense
magnetic field of the cyclotron and the phases of the associated wave
function are shifted, leading to an observable phase shift \cite{SR}. For a
cyclotron of standard size, SR show that the limit 
\begin{equation}
m_{\gamma }^{-1}=10^{6}\,m_{\gamma BD}^{-1}\simeq 2\times 10^{13}cm  \notag
\end{equation}%
is achievable. With their table-top experiment, BD obtained the value $%
m_{\gamma BD}^{-1}\simeq 140Km$ that is equivalent to $m_{phBD}=2.5\times
10^{-45}g$. With SR approach, the new limit of the photon mass is $%
m_{ph}\simeq 2\times 10^{-51}g$ which is of the same order of magnitude of
that found by Luo \textit{et al.} \cite{lu}. Of course, by increasing the
size of the cyclotron a better limit could be obtained. With the standard
technology available, we expect that the limit $m_{ph}\simeq 2\times
10^{-52}g$ is not out of reach.

\subsection{The scalar Aharonov-Bohm effect and the photon mass}

Having exploited the magnetic AB effect in the previous section, we consider
now the scalar AB effect. In this effect charged particles interact with an
external scalar potential $V$. The standard phase $\varphi _{s}$ acquired
during the time of interaction is $\varphi _{s}=\frac{1}{\hbar }\int
eV\left( t\right) dt$.

In the actual test of the scalar AB effect, a conducting cylinder of radius $%
R$ is set at the potential $V$ during a time $\tau $ while electrons travel
inside it. Since no forces act on the charges it is a field-free quantum
effect. If the photon mass does not vanish the potential is modified
according to Proca equation. Gauss' law is modified and the potential $\Phi $
obeys the equation $\nabla ^{2}\Phi -m_{\gamma }^{2}\Phi =0$, with the
boundary condition that the potential on the cylinder be $V$. In cylindrical
coordinates the solutions are the modified Bessel functions of zero order, $%
I_{0}\left( m_{\gamma }\rho \right) $ and $K_{0}\left( m_{\gamma }\rho
\right) $ which are regular at the origin and infinite, respectively. It
follows that the acceptable solution is

\begin{equation}
\Phi \left( \rho \right) \simeq V\left[ 1+\frac{m_{\gamma }^{2}}{2}\left(
\rho ^{2}-R^{2}\right) \right]  \label{2}
\end{equation}%
where the first two terms of the expansion of $I_{0}\left( m_{\gamma }\rho
\right) $ have been considered \cite{Neyenhuis-Mass-Photon}.

For two interfering beams of charges passing through separate cylinders, the
relative phase shift is

\begin{equation}
\delta \varphi _{s}=\frac{1}{\hbar }\int e\left[ V_{1}\left( t\right)
-V_{2}\left( t\right) \right] dt  \label{3}
\end{equation}%
where $V_{1}\left( t\right) $ and $V_{2}\left( t\right) $ are the potentials
applied to cylinder 1 and 2, respectively. Consequently, according to (\ref%
{2}), the contribution of the photon mass to the relative phase shift is

\begin{equation}
\delta \varphi =\delta \varphi _{s}+\Delta \varphi =\delta \varphi _{s}+%
\frac{m_{\gamma }^{2}}{4}\left( \rho ^{2}-R^{2}\right) \delta \varphi _{s}.
\label{4}
\end{equation}%
Obviously, this additional phase shift term vanishes if $m_{\gamma }$
vanishes and the standard result is recovered. The last term of (\ref{4}) is
useful for determining the photon mass in a table-top experiment. We
consider the simple case of one beam travelling inside cylinder 1 and the
other travelling outside it ($V_{2}\left( t\right) =0$) for a short time
interval $\tau $. It follows that $\Delta \varphi =\delta \varphi -\delta
\varphi _{s}$ reads

\begin{equation}
\Delta \varphi =-\frac{em_{\gamma }^{2}}{4}\left( \rho ^{2}-R^{2}\right) V%
\frac{\tau }{\hbar }  \label{5}
\end{equation}%
where $V=V_{1}\left( t\right) -V_{2}\left( t\right) $. This is our main
result for determining the photon mass limit. Interferometric experiments
may be performed with a precision of up to $10^{-4},$ therefore, following
the approaches of BD and SR we set $\Delta \varphi =\varepsilon $, $%
\varepsilon =10^{-4}$. Also, we suppose that the beam 1 travels nearly at
the centre of the cylinder ($\rho \ll R$) so that

\begin{equation}
m_{\gamma }^{-1}=\frac{R}{2}\sqrt{\frac{\pi V\tau }{\varepsilon (h/2e)}}
\label{6}
\end{equation}%
The following values may be used to estimate $m_{\gamma }^{-1}$: $V=10^{7}V$%
, $h/2e=2.067\times 10^{-15}Tm^{2}$, $\tau =5\times 10^{-2}s$ and $R=27cm$.
The corresponding range of the photon mass is%
\begin{equation}
m_{\gamma }^{-1}=3,4\times 10^{13}cm  \label{mm1}
\end{equation}%
which yields the improved photon mass limit $m_{ph}=9,4\times 10^{-52}g$,
but we are left to justify the values used above for $\tau $ and $R$, which
are both quite high.

It is interesting to compare the strength of the AB phase of the scalar AB
effect with that of the magnetic AB effect. The scalar AB phase may be
expressed as $eV$ $\tau /\hbar $, while the magnetic AB phase is $%
eAL/(c\hbar )$, and the link between the particle's classical path is $%
L=\tau v$ with $v$ its speed assumed to be uniform. According to special
relativity, magnetism is a second order effect of electricity, therefore in
normal conditions the strength of the coupling $eA/c$ is smaller than the
coupling $eV$. As a consequence of this, the phase variation due to the
finite photon mass should be smaller in the magnetic than in the scalar AB
effect. In other words, the scalar AB effect should be yielding a better
limit for the photon mass than the magnetic AB effect. However, the above
consideration is valid if in the actual experiments we have comparable path
lengths, i.e., if $\tau \simeq L/v$. In the table-top experiment by SR \cite%
{SR} $L$ is of the order of several meters. Choosing as charged particles
heavy ions, for example $^{133}Cs^{+},$ their speed could be $27m/s$ \cite%
{LVIS}. With this speed and $L=1.35m$ for the cylinder length, we get $\tau
=5\times 10^{-2}s$ for the time of flight inside the cylinder. Since $\tau
\simeq L/v$, the improved result (\ref{mm1}) obtained by exploiting the
scalar AB effect is justified.

However, the high values chosen for $R$ and $L$ imply that the charged
particle beams will have to keep their state of coherence through an
extended region of space $L=1.35m$ during the interferometric measurement
process, while in standard interferometry the path separations are of the
order of at most a few $cm$. Thus, technological advances are needed in this
respect, as also mentioned in the article by SR \cite{SR} and the references
cited therein.

Nevertheless, the feasibility of testing the photon mass with the scalar AB
effect has been confirmed by the recent work of Neyenhuis, Christensen, and
Durfee \cite{Neyenhuis-Mass-Photon}, lending support to the quantum approach.

Actually, it is conceivable the possibility of extending to the case of the
scalar AB effect the techniques of Refs. \cite{sang} and \cite{dow} for a
coherent superposition of beams of charged particles with opposite charge
state $\pm q$, as suggested by SR in Ref. \cite{SR}. This may lead to
achieve even better limits for the photon mass. In fact, by means of these
techniques it is feasible to suppose that the particles paths may be $10^{2}$
times those considered above. Thus, the time of flight $\tau $ becomes $%
10^{2}$ times bigger. Although the technical details will be given
elsewhere, we anticipate that further improvement can be achieved by bending
the particle path into a circular one, as in the case of ions in a
cyclotron. In this case, $\tau $ may be increased $10^{4}$ times. Thus , we
project that a photon mass limit of the order of%
\begin{equation*}
m_{ph}\simeq 10^{-54}g
\end{equation*}
may be achieved.

\section{Conclusions}

We have recalled that the interaction momenta $\mathbf{Q}$ of the effects of
the AB type and of light in moving media have the same physical origin,
i.e., are given by the variation of the momentum of the interaction em
fields $\mathbf{P}_{e}$. Expecting that the effectiveness of the light delay
mechanism in a rarefied gas differs from that of a compact transparent fluid
or solid, we consider a tentative model of light propagation that validates
the analysis made by Consoli et al. \cite{c} and Guerra et al. \cite{4}. As
a test of the speed of light in moving rarefied media and of the preferred
frame velocity, we propose an improved first order optical experiment that
is a variant of the historical Mascart-Jamine experiment.

Finally, we have considered the table-top approach of Boulware and Deser to
the photon mass and verified its applicability to other effects of the AB
type, concluding that the new effect using beams of charged particles with
opposite charge state $\pm q$ for the magnetic AB effect, and the scalar AB
effect are a good candidates for determining the limit of the photon mass.
Using a quantum approach to evaluate the limit of $m_{ph}$ with these
effects, and supposing that the recent interferometric techniques can be
used, we project that a bench-top experiment may yield the limit $%
m_{ph}\simeq 10^{-54}g$, an important result that would improve the limits
achieved with recent classical and quantum approaches. In any event,
advances in this area indicate that quantum approaches the photon mass limit
are feasible and may compete with or surpass the traditional classical
methods.

\section{Acknowledgments}

This work was supported in part by the CDCHT (Project C-1413-06-05-A), ULA, M%
\'{e}rida, Venezuela.


\begin{thebibliography}{99}
\bibitem{sg} G. Spavieri and G. T. Gillies, Chin. J. Phys., \textbf{45}
(2007) 12.

\bibitem{han} J. H. Hannay, unpubl., Cambridge Univ. Hamilton prize essay
(1976); R. J. Cook, H. Fearn, and P. W. Milonni, Am. J. Phys. \textbf{63}
(1995) 705.

\bibitem{AB} Y. Aharonov and D. Bohm, Phys. Rev. \textbf{115} (1959) 485; Y.
Aharonov and A. Casher, Phys. Rev. Lett. \textbf{53} (1984) 319; G.
Spavieri, Phys. Rev. Lett. \textbf{81} (1998) 1533, Phys. Rev. A \textbf{59}
(1999) 3194; V. M. Tkachuk, Phys. Rev. A \textbf{62} (2000) 052112-1.

\bibitem{s} G. Spavieri, Phys. Rev. Lett. \textbf{82}, 3932 (1999); Phys.
Lett. A, \textbf{310}, 13 (2003); Eur. J. Phys. D, \textbf{37} (2006) 327.

\bibitem{sgo} G. Spavieri and G. T. Gillies, Nuovo Cimento, \textbf{118} B
(2003) 205; G. Spavieri, L. Nieves, M. Rodriguez, and G. T. Gillies. \textit{%
Has the last word been said on Classical Electrodynamics?-New Horizons},
Rinton Press, USA (2004) 255.

\bibitem{sg1} G. Spavieri, G. T. Gillies et al., in \textit{Ether, Spacetime
\& Cosmology}, (2009) in press.

\bibitem{1} J. S. Bell, Speakable and Unspeakable in Quantum Mechanics,
Cambridge Univ. Press, Cambridge, 1988; C. Leubner, K. Aufinger, P. Krumm,
Eur. J. Phys. \textbf{13} (1992) 170. F. Selleri, Found. Phys. \textbf{26}
(1996) 641; Found. Phys. Lett. \textbf{18} (2005) 325.

\bibitem{4} R. de Abreu, V. Guerra, Relativity--Einstein's Lost Frame, 1st
ed., Extra]muros[, Lisboa, 2005. V. Guerra and R. de Abreu, Found. Phys. 
\textbf{36} (200691826; V. Guerra and R. de Abreu, Phys. Lett. A \textbf{333}
(2004) 355.

\bibitem{c} M. Consoli, E. Costanzo, Phys. Lett. A \textbf{333} (2004) 355;
astro-ph/0311576; M. Consoli, A. Pagano and L. Pappalardo, Phys. Lett. A%
\textbf{\ 318} (2003) 292; M. Consoli, Phys. Rev. D \textbf{65} (2002)
105017; Phys. Lett. B \textbf{541} (2002) 307; M. Consoli and E. Costanzo,
Phys. Lett. A \textbf{361} (2007) 513.

\bibitem{ck} R.T. Cahill, K. Kitto, physics/0205070; Apeiron \textbf{10}
(2003) 104; R.T. Cahill, Apeiron \textbf{11} (2004) 53.

\bibitem{mil} D.C. Miller, Rev. Mod. Phys. \textbf{5}, 203 (1933) .

\bibitem{r9} L. Indorato and G. Masotto, Annals of Science, \textbf{46},
117-163 (1989).

\bibitem{Sem} G. Spavieri, Eur. Phys. J. D, \textbf{39}, 157 (2006).

\bibitem{fre} A. J. Fresnel, Ann. Chim. (Phys.) \textbf{9,} 57 (1818). H.
Fizeau, C. R. Acad. Sci. (Paris) \textbf{33}, 349 (1851).

\bibitem{boy} See: T. H. Boyer, Phys. Rev. D \textbf{8} (1973) 1667; X. Zhu
and W. C. Henneberger, J. Phys. A \textbf{23} (1990) 3983; G. Spavieri, in
Refs. \cite{s}.

\bibitem{duf} M. Duffy, private comm., Int. Conf. \textit{Physical
Interpretation of Relativity Theory} 2006.

\bibitem{tan} F. R. Tangherlini, Supp. Nuovo Cimento \textbf{20 }(1961) 1;
T. Sjodin, Nuovo Cimento, B \textbf{51} (1979) 299; T. Sjodin and M. F.
Podlaha, Lett. Nuovo Cimento, \textbf{31} (1982) 433; R. Mansouri and R. V.
Sexl, Gen. Rel. Grav., \textbf{8}, (1977) 497, 515, 809.

\bibitem{sgv} G. Spavieri, G. T. Gillies, V. Guerra, and R. De Abreu, EPJD
in press (2008).

\bibitem{mj} E. Mascart and J. Jamine, Ann. \'{E}c. norm. \textbf{3} (1874)
336.

\bibitem{wfh} E. R. Williams, J. E. Faller and H. A. Hill, Phys. Rev. Lett., 
\textbf{26}, 721 (1971); L. Davis, A.S. Goldhaber and M. M. Nieto, Phys.
Rev. Lett., \textbf{35}, 1402 (1975); P. A. Franken and G. W. Ampulski,
Phys. Rev. Lett, \textbf{26}, 115 (1971); J. J. Ryan, F. Accetta, and R. H.
Austin, Phys. Rev. D, \textbf{32}, 802 (1985); R. Lakes, Phys. Rev. Lett. 
\textbf{80}, 1826 (1998).

\bibitem{lu} J. Luo, L.-C. Tu, Z. K. Hu, and E.-J. Luan, Phys. Rev. Lett., 
\textbf{90}, 081801-1 (2003); L.-C. Tu, J. Luo, and G. T. Gillies, Rep.
Prog. Phys. \textbf{68}, 77 (2005).

\bibitem{bd} D. G. Boulware and S. Deser, Phys. Rev. Lett., \textbf{63},
2319 (1989).

\bibitem{sdual} G. Spavieri, Phys. Lett. A, \textbf{310}, 13 (2003).

\bibitem{sep} G. Spavieri, Eur. J. Phys. D, \textbf{37}, 327 (2006).

\bibitem{sang} K. Sangster, E.A. Hinds, S.M. Barnett, E. Riis, Phys. Rev.
Lett. \textbf{71}, 3641 (1993); K. Sangster, E.A. Hinds, S. M. Barnett, E.
Riis, A.G. Sinclair, Phys. Rev. A \textbf{51}, 1776 (1995); see also R.C.
Casella, Phys. Rev. Lett. \textbf{65}, 2217 (1990).

\bibitem{dow} J.P. Dowling, C.P. Williams, J.D. Franson, Phys. Rev. Lett. 
\textbf{83}, 2486 (1999).

\bibitem{SR} G. Spavieri and M. Rodriguez, Phys. Rev. A \textbf{75}, 052113
(2007).

\bibitem{Neyenhuis-Mass-Photon} B. Neyenhuis, D. Christensen, D. S. Durfee,
Phys. Rev. Lett. \textbf{99}, 200401 (2007).

\bibitem{LVIS} Z. T. Lu, K. L. Corwin, M. J. Renn, M. H. Anderson, E. A.
Cornell, and C. E. Wieman, Phys. Rev. Lett. \textbf{77}, 3331 (1996).
\end{thebibliography}
\end{document}